\documentclass[aps, preprint, superscriptaddress, showkeys]{revtex4-2}
\usepackage{amssymb, amsmath}
\usepackage{xr}
\usepackage{hyperref}
\hypersetup{colorlinks=true, citecolor=blue, urlcolor=blue, linkcolor=blue}

\usepackage{graphicx}
\usepackage{soul}
\usepackage{xcolor}
\usepackage[hmargin=2cm,vmargin=2.5cm]{geometry}
\usepackage{filecontents}

\newcommand{\ens}{Laboratoire de Physique de l'Ecole normale sup\'erieure, ENS, Universit\'e PSL, CNRS, Sorbonne Universit\'e, Universit\'e de Paris, 24 rue Lhomond, 75005 Paris, France}
\newcommand{\trt}{Thales Research \& Technology, 91767 Palaiseau, France}
\newcommand{\iuf}{Institut universitaire de France}

\begin{document}

\title{Raman spectroscopy of monolayer to bulk \texorpdfstring{PtSe\textsubscript{2}}{PtSe2} exfoliated crystals}

\author{Marin Tharrault}
\email{marin.tharrault@phys.ens.fr}
\affiliation{\ens}
\author{Eva Desgu\'e}
\affiliation{\trt}
\author{Dominique Carisetti}
\affiliation{\trt}
\author{Bernard Plaçais}
\affiliation{\ens}
\author{Christophe Voisin}
\affiliation{\ens}
\author{Pierre Legagneux}
\affiliation{\trt}
\author{Emmanuel Baudin}
\email{emmanuel.baudin@phys.ens.fr}
\affiliation{\ens}
\affiliation{\iuf
\vspace{8ex}
}

\keywords{2D materials, TMD, PtSe2, Raman spectroscopy, intrinsic properties, exfoliation, quality metrics}

\begin{abstract}
Raman spectroscopy is widely used to assess the quality of 2D materials thin films.
This report focuses on $\rm{PtSe_2}$, a noble transition metal dichalcogenide which has the remarkable property to transit from a semi-conductor to a semi-metal with increasing layer number. 
While polycrystalline $\rm{PtSe_2}$ can be grown with various crystalline qualities, getting insight into the monocrystalline intrinsic properties remains challenging. 
We report on the study of exfoliated 1 to 10 layers $\rm{PtSe_2}$ by Raman spectroscopy, featuring record linewidth. The clear Raman signatures allow layer-thickness identification and provides a reference metrics to assess crystal quality of grown films.
\end{abstract}

\maketitle

\newpage

Transition Metal Dichalcogenides (TMDs) are promising materials for future electronic and optoelectronic devices, owing to their large optical absorption per layer and high electronic mobility \cite{Zhao2016SC_SM, Bonell2022MBE}. Furthermore, they feature strong van der Waals interlayer coupling, resulting in a tunable layer-dependent band structure. Among TMDs, thin films of Platinum Diselenide ($\rm{1T-PtSe_2}$) are semi-conductors and feature exceptional bandgap variations, with a transition to a semi-metal with increasing thickness \cite{Zhao2016SC_SM, Yu2018NtuNatCom, Villaos2019DFT, Kandemir2018RamanDFT}.
For this reason it can reach small bandgap values, permitting efficient photodetection in the infrared range \cite{Zhao2016SC_SM, Yu2018NtuNatCom, Ma2021ElectroExfoPhotodetect}. This makes $\rm{PtSe_2}$ a promising building block for optoelectronic devices operating in the telecom band, and several growth methods (CVD, TAC, MBE) are being developed to provide high-quality scalable films for industry \cite{Yu2018NtuNatCom, Wang2016NanoSheet, Jiang2019TAC, Yan2017MBE, Han2019CvdHorz, Hilse2020Selenization, Bonell2022MBE, articlePtSe2absorption}.
These films are commonly characterized using a variety of methods, including electron diffraction, X-ray spectroscopy or diffraction, electron microscopy, atomic force microscopy, electronic transport measurements and optical spectroscopy \cite{Gatensby2014RamanTemXps, Bonell2022MBE}.

Among these methods, Raman spectroscopy presents several key advantages: fast, cheap and non-contact, it probes optical phonon transitions -- highly sensitive to defects -- and is therefore used as a primary characterization to identify the structure and assess film quality \cite{Neumann20152DRamanPeak, Banszerus2017RamanSubstrate, Mignuzzi2015RamanIrradMoS2, Pierce2018IrradGrMoS2Raman, Stenger2017RamanHBn}.
Previous works on $\rm{PtSe_2}$ established the most salient Raman spectral features \cite{OBrien2016Raman, Szydowska2020RamanLPE, Gulo2020TempOpticalRaman, Lukas2021TAC, Yin2021RamanTemp, Yasuda2023RamanHelicity, Raczynski2023TempRaman}: they identified the optical vibrational phonon modes associated with the peaks, detailed the Raman peaks intensities and shifts evolution with the thickness, proposed quality metrics, and studied the polarization and temperature dependencies of the Raman peaks.
However, due to the limited quality of the studied samples, the precise features of the Raman signature of few-layer intrinsic $\rm{PtSe_2}$ remained out of reach.

In this work, we report high-resolution Raman spectroscopy of record-quality exfoliated $\rm{PtSe_2}$ crystals of layer-controlled thickness.
We establish criteria for film quality, and demonstrate that the characteristic Raman signature of thin flakes enables the determination of layer count.
By providing reference Raman spectra of exfoliated single crystals, this work will guide the development of the continuously improving $\rm{PtSe_2}$ growth technology \cite{Jiang2019TAC, Bonell2022MBE, articlePtSe2absorption}.

\section*{Samples Fabrication}

\begin{figure*}[h!]
\begin{center}
\includegraphics[width=6.75in]{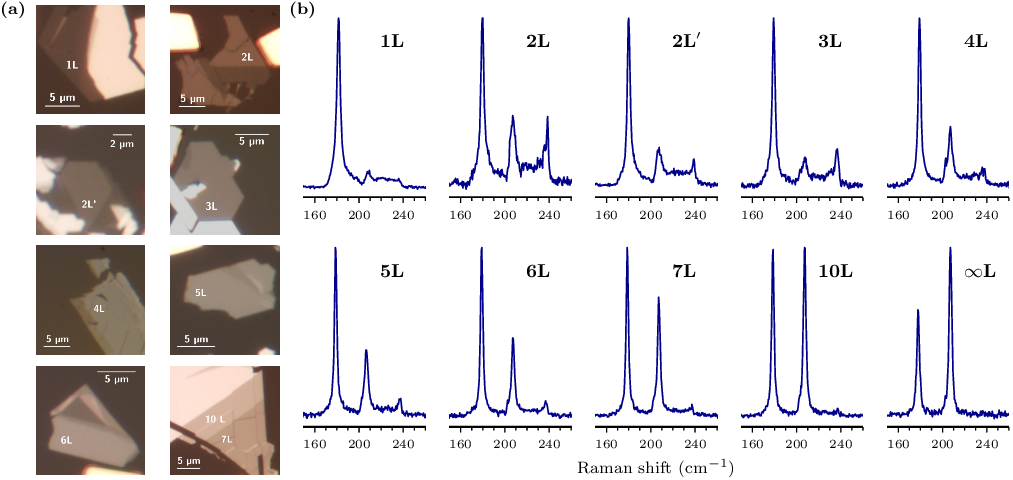}
\end{center}
\caption{$\rm{PtSe_2}$ exfoliated flakes and their Raman spectra. (a) Micrographs of $\rm{PtSe_2}$ flakes, each region of interest is labeled with its layer count. (b) The associated Raman spectra displayed on the $150\,-\,260\,\rm{cm^{-1}}$ range, and normalized to their maximum amplitude.
  \label{fig:ramanSpectraPicsIdNLayers}}
\end{figure*}

Chemical Vapour Transport grown $\rm{PtSe_2}$ crystals (HQ Graphene \cite{HQGraphenePtSe2}) are exfoliated on fused silica substrates, as shown in \textbf{figure \ref{fig:ramanSpectraPicsIdNLayers}a}. Au-assisted mechanical exfoliation provides flakes as thin as a monolayer thanks to the strong affinity of gold with Selenium atoms \cite{Desdai2016AuExfo, Huang2020AuExfo, Liu2020AuExfo}.

Each flake number of layers is identified using absorption microspectroscopy and confirmed by Atomic Force Microscopy, as described in detail in reference \cite{articlePtSe2absorption}. 

\section*{Atomic structure signature} 

High-resolution Raman spectroscopy is performed using a $50\,\rm{\mu W}$ low-power $514\,\rm{nm}$ green laser source, together with a $50\,\rm{cm^{-1}}$ edgepass filter (additional details in experimental section).

Spectra of selected flakes are displayed in figure \ref{fig:ramanSpectraPicsIdNLayers}b and consist in 3 solitary peaks in the $150\,-\,260\,\rm{cm^{-1}}$ spectral range. The flake thickness has little influence on the total scattered Raman intensity. This can be understood by the fact that green light is resonantly absorbed regardless of the number of layers \cite{articlePtSe2absorption}.
The full set of spectra is included in the Supplementary Material (SM), figure \ref{fig:ramanSpectraFull}.

\begin{figure*}[h!]
\begin{center}
\includegraphics[width=6.75in]{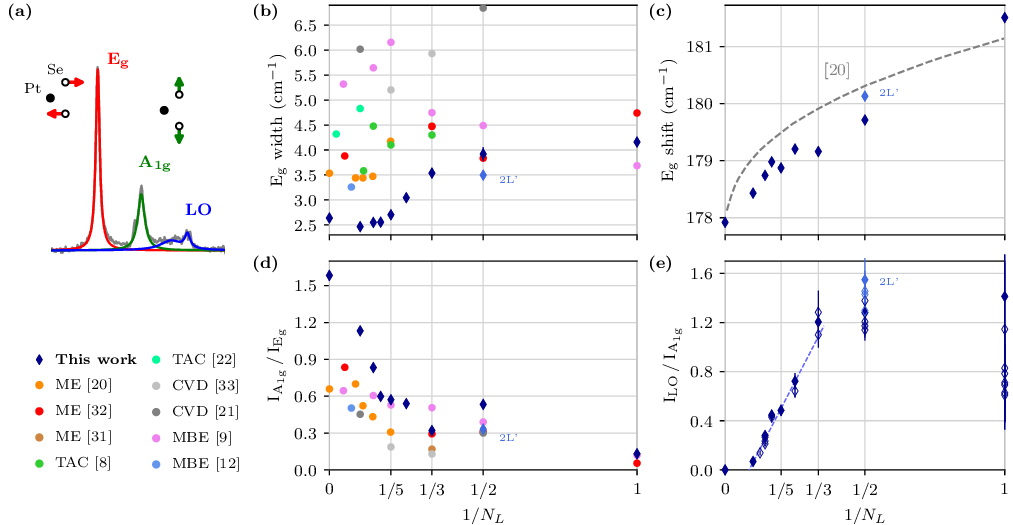}
\end{center}
\caption{Evolution of Raman spectra characteristics with layer count and film quality. (a) Decomposition of the 4-components Lorentzian fit into three contributions : $E_g$, $A_{1g}$ and $LO$ modes, and schematic of atomic motions for $E_g$ and $A_{1g}$ modes. (b) $E_g$ peak width and (c) shift (the dashed line is a law derived from different studies in reference \cite{Szydowska2020RamanLPE}). (d) Intensities ratio of $A_{1g}$ to $E_g$ modes and (e) $LO$ to $A_{1g}$ modes. Errorbars are included, displaying one standard deviation fitting uncertainty. The empty diamonds incorporate data not shown in figure \ref{fig:ramanSpectraPicsIdNLayers}, and the dashed line is a linear fit for $3$ to $10$ layers. For the sake of comparison, in (b) and (d) we fitted digitalized Raman data from the literature \cite{Szydowska2020RamanLPE, Das2021TransferedContacts3L, Bae2021PumpProbeExfo, Jiang2019TAC, Lukas2021TAC, Xu2019TransportDC, Gulo2020TempOpticalRaman, Yan2017MBE, articlePtSe2absorption}, performed with green $532\,\rm{nm}$ or $514\,\rm{nm}$ light and where layer count is determined from film thickness considering an individual layer thickness of $0.5\,\rm{nm}$ \cite{articlePtSe2absorption}. 2L' is displayed in lighter blue to highlight the difference with other samples. 
\label{fig:ramanParameters}}
\end{figure*}

The spectral shapes correspond to the 1T phase of $\rm{PtSe_2}$, identified with its 4 optical phonon modes: the Raman-active $E_g$ and $A_{1g}$ modes around $180\,\rm{cm^{-1}}$ and $210\,\rm{cm^{-1}}$ respectively, and the IR-active $A_{2u}$ and $E_u$ modes gathered as $LO$ (longitudinal optical) around $230\,\rm{cm^{-1}}$\cite{OBrien2016Raman, Kandemir2018RamanDFT, Fang2019AA_AB_DFT}, as depicted in \textbf{figure \ref{fig:ramanParameters}a}.

Raman modes spectral weights and widths are assessed by a 4-components Lorentzian fit.
Each Lorenzian function $L(\tilde{\nu})$ is parameterized by its shift $\tilde{\nu}_0$, its integrated intensity $I$, and full width at half maximum (thereafter abbreviated as width) $\Gamma$, such that:

\begin{equation}
L(\tilde{\nu}) = \frac{I}{\pi}\frac{\Gamma/2}{(\tilde{\nu}-\tilde{\nu}_0)^2 + (\Gamma / 2)^2}
\end{equation}

Selected fit parameters are gathered in figures \ref{fig:ramanParameters}b-e. More extensive data can be found in figure SM\ref{fig:ramanFullParameters}.  Furthermore, the Raman signatures are mostly uniform over the flakes surfaces (down to the measurement accuracy, SM\ref{fig:ramanMaps}).

Almost all samples are identified using optical absorption spectroscopy as the most-stable AA stacking, as their optical absorption spectra match Density Functional 
Theory (DFT) predictions \cite{articlePtSe2absorption}. 2L' samples are a noticeable exception, with a singular Raman signature differing from the one of 2L (detailed comparison in SM\ref{fig:ramanBilayers}). In fact, the comparison of absorption behavior with DFT predictions for 2L' suggests an AB stacking \cite{articlePtSe2absorption}, which has been theoretically identified as a stable phase for $\rm{1T-PtSe_2}$ \cite{Fang2019AA_AB_DFT, Kandemir2018RamanDFT, Kempt2022Stackings, articlePtSe2absorption}, and is a common defect in grown films \cite{Xu2021STEMstackAB, Ryu2019STEMstackAB}.

The influence of the substrate on the Raman signature is moreover investigated. We compare the Raman spectra of few-layers flakes deposited on fused silica and crystalline sapphire (figure SM\ref{fig:ramanSapphire}). No noticeable difference can be observed, the Raman signature appearing to be identical. It is likely due to the fact that in both cases, the flakes and the substrate are not directly in van der Waals contact, due to the presence of interfacial layers of water \cite{Rokni2020WaterLayer}.

\section*{Crystalline Quality}

In TMDs and graphene, Raman mode linewidth is very sensitive to the defect level
 \cite{Neumann20152DRamanPeak, Banszerus2017RamanSubstrate, Mignuzzi2015RamanIrradMoS2, Pierce2018IrradGrMoS2Raman}. $\rm{PtSe_2}$ is no exception, and the $E_g$ linewidth has been shown to increase with film defectiveness \cite{Lukas2021TAC, Szydowska2020RamanLPE, Jiang2019TAC}. 
For the 1L flakes studied in this work, a narrow $E_g$ peak is associated with the presence of sharp optical features in the absorption spectrum \cite{articlePtSe2absorption}, again indicating the relevance of this Raman spectral signature.

Recent works reported the growth of good quality $\rm{PtSe_2}$ films, using several methods: Thermally Assisted Conversion (TAC) \cite{Jiang2019TAC, Lukas2021TAC}, Chemical Vapor Deposition (CVD) \cite{Xu2019TransportDC, Gulo2020TempOpticalRaman} and Molecular Beam Epitaxy (MBE) \cite{Yan2017MBE, articlePtSe2absorption}. Some research focused instead on Mechanichally Exfoliated (ME) $\rm{PtSe_2}$ crystals \cite{Szydowska2020RamanLPE, Das2021TransferedContacts3L, Bae2021PumpProbeExfo}. These works stand out for their narrow $E_g$ peaks (figure \ref{fig:ramanParameters}b), with linewidths values below $6\,\rm{cm^{-1}}$ -- reaching about $3.5\,\rm{cm^{-1}}$ for thick ME, TAC and MBE samples, about $4\,\rm{cm^{-1}}$ for ME and MBE 1L and 2L samples (however the film continuity of the MBE 1L is not established). 

In our work, the high quality of the $\rm{PtSe_2}$ flakes is assessed by the unprecedented narrow linewidth of this $E_g$ mode, from monolayer to bulk thicknesses. The $E_g$ linewidth reaches $4.2\,\rm{cm^{-1}}$ for $\rm{1L}$, $3.9\,\rm{cm^{-1}}$ for $\rm{2L}$ and around $2.5\,\rm{cm^{-1}}$ for thick samples ($\rm{7L}$ to bulk).
Among the $\rm{1L}$ exfoliated flakes studied, several feature large $E_g$ linewidths ($> 6\,\rm{cm^{-1}}$), as shown in figure SM\ref{fig:ramanFullParameters}. We could not identify if such flakes were damaged during the exfoliation process, or if they originated from more defective areas of the original CVT crystal.
As reported by the aforementioned works, the $E_g$ peak features of strong increase of its linewidth with decreasing film thickness (figure \ref{fig:ramanParameters}b). This increase is commonly attributed to increasing defectiveness as the film thickness diminishes \cite{OBrien2016Raman}.

\begin{figure*}[h!]
\begin{center}
\includegraphics[width=3.375in]{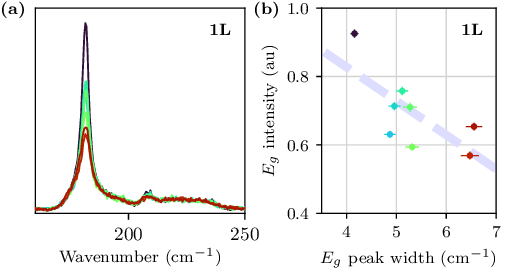}
\end{center}
\caption{$E_g$ peak variability of monolayer samples. (a) Raman spectra, and (b) their $E_g$ peak intensity versus width -- the trend is given by a linear fit (dashed line). In both figures, the color represents the measured $E_g$ linewidth.
\label{fig:EgMode}}
\end{figure*}

Let us now turn our examination from the $E_g$ peak width to its absolute intensity, which appears as a good criterion for film quality. 
Indeed, for a given layer count, the thin exfoliated $\rm{PtSe_2}$ flakes feature a large sample-to-sample variability of their $E_g$ peak integrated intensities (figure \ref{fig:EgMode}a for the 1L case), while the $A_{1g}$ and $LO$ modes are unaffected.
We observe that higher $E_g$ peak intensity correlates with narrower linewidth (figure \ref{fig:EgMode}b, more thicknesses are studied in figure SM\ref{fig:EgWidthInt}), therefore appearing as a robust metric for film quality. 
This sensitivity of the $E_g$ mode intensity to flake quality  may be due to the high sensitivity to the electromagnetic environment and disorder of Fröhlich electron-phonon coupling which affects $E$ longitudinal optical modes amplitudes in polar crystals \cite{YuCardona2010SC, Miller2019FrohlichExciton}.

\section*{Number of Layers Identification}

It is tempting to identify the number of layers from Raman signatures, in particular in $\rm{PtSe_2}$ in which the strong thickness-dependence Raman spectra led several authors to propose metrics relying on the shift of the $E_g$ peak or the ratio of $A_{1g}$ to $E_g$ peak intensity. In this section, we show that these methods are imprecise, and we propose a new method for exfoliated $\rm{PtSe_2}$ crystals. 

The $E_g$ peak position exhibits a quasi-linear blueshift with the inverse number of layers $1\,/\,N_L$. The observed dependency in figure \ref{fig:ramanParameters}c is qualitatively compatible with previous studies, compiled as a grey dashed line and proposed as a metric by reference \cite{Szydowska2020RamanLPE}. However, the discrepancy between this metrics and our data is too large to allow layer count identification.

The $A_{1g}$ to $E_{g}$ peaks intensities is commonly considered as characteristic of the layer count \cite{Shi2019CvdAu, Szydowska2020RamanLPE, Qiu2021PumpProbe}.
Indeed the $A_{1g}$ mode involves out-of-plane motion of external $\rm{Se}$ atoms (figure \ref{fig:ramanParameters}a) and its intensity rises with the thickness, while longitudinal optical $LO$ modes disappear for high layer number \cite{OBrien2016Raman}. These trends are observed as well in our work (figure  \ref{fig:ramanSpectraPicsIdNLayers}b), but appear to be more complex for 1L and 2L. 
Nonetheless, the $A_{1g}$ to $E_g$ intensities ratio does not appear as a reliable signature of the layer count, as it differs significantly from one research to another (figure \ref{fig:ramanParameters}d, references therein). 
Moreover, as described above, we observe for low layer count important variations of the $E_g$ peak intensity resulting in $I_{A_{1g}}/I_{E_g}$ dispersion. 

We propose instead to identify the number of layers by using the $A_{1g}$ and $LO$ peaks which are more robust to the defect level. The $A_{1g}$ to $LO$ intensity ratio for each sample is represented in figure \ref{fig:ramanParameters}e, as a function of inverse number of layers $1/N_L$. One can see straight away  that the thickness can be evaluated from $3$ to $10$ layers using the linear law displayed in figure \ref{fig:ramanParameters}e:

\begin{equation}
    N_L = \frac{4.38}{I_{LO}\,/\,I_{A_{1g}} + 0.38}
\end{equation} 

Mono- and bi-layers are exceptions to this law, but can be easily identified. Both present similar $A_{1g}$ and $LO$ peaks intensities, and 1L has a remarkably low $A_{1g}$  to $E_g$ peaks intensities ratio, with $I_{A_{1g}}/I_{E_g} < 35\,\%$  with a singularly wide $LO$ peak, while 2L has relatively high ratio $I_{A_{1g}}/I_{E_g} > 45\,\%$, and a sharper $LO$ peak (figure SM\ref{fig:ramanFullParameters}).

Beyond $10$ layers, accurate determination of the layer count isn't critical anymore as AFM can provide a fair estimate.

\vspace{2ex}

\section*{Conclusion}

By using exfoliated CVT $\rm{PtSe_2}$ crystals with layer-defined thickness, this study provides reference Raman spectrographs of high-quality flakes, which are particularly important for benchmarking emerging high-quality growth methods.
We showed that crystalline quality can be assessed from the width and height of the $E_g$ mode peak, with $E_g$ linewidth narrowing down to $4.2\,\rm{cm^{-1}}$ for monolayer and $2.5\,\rm{cm^{-1}}$ for thicker exfoliated $\rm{PtSe_2}$. We observed that commonplace criteria for layer count identification, either based on $E_g$ peak shift or on $A_{1g}$ to $E_g$ peaks intensities ratio, leaves much to be desired for exfoliated high-quality $\rm{PtSe_2}$. This led us to propose a robust method in this latter case based on the $A_{1g}$ and $LO$ specific peaks pattern.

\section*{Experimental Methods}

\subsection*{Samples}
A $60\,\rm{nm}$ gold film is deposited using evaporation on a fused silica substrate (QX/QS Quartz Suprasil 300 from Hellma Analytics), on which the crystals are pre-exfoliated. The samples are then annealed at $150\rm{^\circ C}$ and the gold film is peeled using thermal release tape, thereby detaching few-layer $\rm{PtSe_2}$. The peeled gold film is then transferred to a target substrate, and the gold is removed using $\rm{KI_2}$ etching.

\subsection*{Raman spectrometry}
We use a Raman Qontor spectrometer from Renishaw with a $3000\,\rm{gr/mm}$ grating, a $\times 100$ microscope objective, a $514\,\rm{nm}$ laser source and a $50\,\rm{cm^{-1}}$ edgepass filter. The laser is operated at a power of $50\,\rm{\mu W}$, in order to suppress any thermal shift or broadening of the $\rm{PtSe_2}$ lines (the spectral shift of the main $E_g$ peak was evaluated to reach approximately $0.1\,\rm{cm^{-1}/m W}$). Each spectrum is integrated for $1\,\rm{min}$ and averaged 5 times. Muon peaks are eliminated from the spectra and the baseline is removed using a 3rd order polynomial fit of the background signal.
The spectrometer is calibrated using a $\mathrm{Si_{100}}$ crystal which possesses a Raman peak centered at $520.45\,\mathrm{{cm^{-1}}}$ \cite{Itoh2020RamanSiLine}.

\section*{Acknowledgements}

The authors acknowledge the financial support from the European Union’s Horizon 2020 program under grant agreement no. 785219, no.  881603 (Core2 and 3 Graphene Flagship), as well as from ANR-2018-CE08-018-05 (BIRDS) and ANR-2021-CE24-0025 (ELuSeM).

\section*{Data availability}

The data that support the findings of this study are openly available, at \cite{Tharrault2023DataRaman}.

\bibliography{ref}

\makeatletter\@input{xx.tex}\makeatother
\end{document}


\title{Raman spectroscopy of monolayer to bulk \texorpdfstring{PtSe\textsubscript{2}}{PtSe2} exfoliated crystals - Supplementary Material}

\author{Marin Tharrault}
\email{marin.tharrault@phys.ens.fr}
\affiliation{\ens}
\author{Eva Desgu\'e}
\affiliation{\trt}
\author{Dominique Carisetti}
\affiliation{\trt}
\author{Bernard Plaçais}
\affiliation{\ens}
\author{Christophe Voisin}
\affiliation{\ens}
\author{Pierre Legagneux}
\affiliation{\trt}
\author{Emmanuel Baudin}
\email{emmanuel.baudin@phys.ens.fr}
\affiliation{\ens}
\affiliation{\iuf}

\maketitle

\begin{figure*}[h!]
\begin{center}
\includegraphics[width=7in]{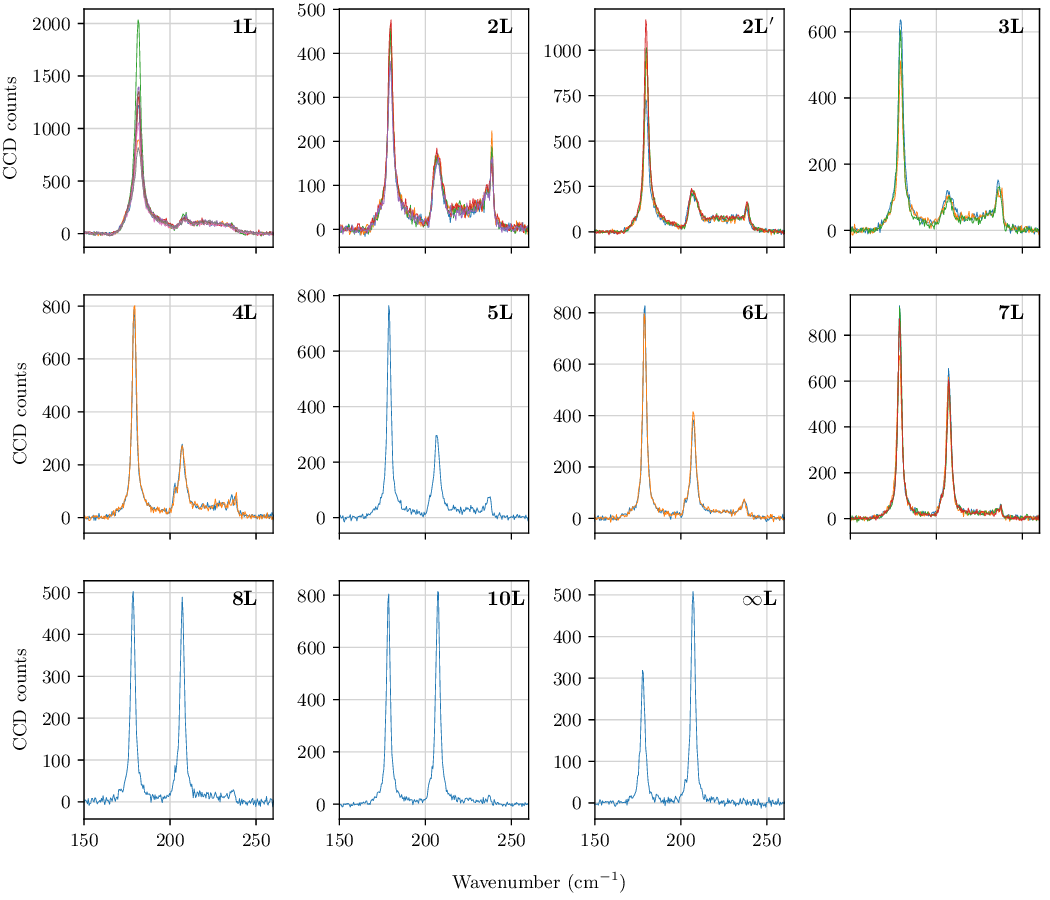}
\end{center}
\caption{Raw Raman spectra where the baseline has been substracted. Each spectrum originates from a different flake (for a total of 32 flakes).
  \label{fig:ramanSpectraFull}}
\end{figure*}

\begin{figure*}[h!]
\begin{center}
\includegraphics[width=6.5in]{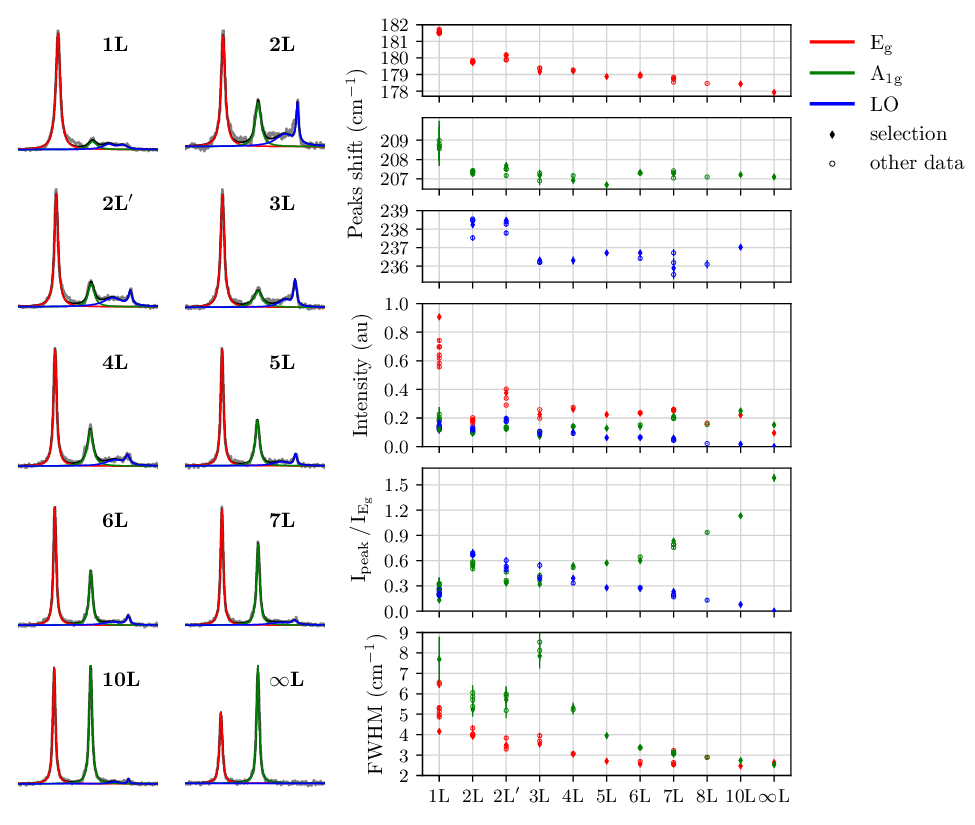}
\end{center}
\caption{Raman spectra peaks parameters. Left panel: selected spectra displayed in main text fig. \ref{fig:ramanSpectraPicsIdNLayers} with the fit decomposition in $\rm{E_{g}}$, $\rm{A_{1g}}$ and $LO$ contributions in respectively red, green and blue. Right panel: The fit parameters extracted for all the measured spectra: peaks shifts, intensities, intensities ratios and full-width at half maximum (the $LO$ peak shift refers to the position of the LO maximum amplitude). The selected spectra parameters are displayed as diamonds, and others as circles. Only well-fitted parameters are represented on these plots. Each datapoint is displaying a Raman mode parameter originating from a different flake. The fit is performed by attributing a $1/\sqrt{A}$ weight to each data point ($A$ being the amplitude), to account for photon shot noise.
  \label{fig:ramanFullParameters}}
\end{figure*}

\begin{figure*}[h!]
\begin{center}
\includegraphics[width=7in]{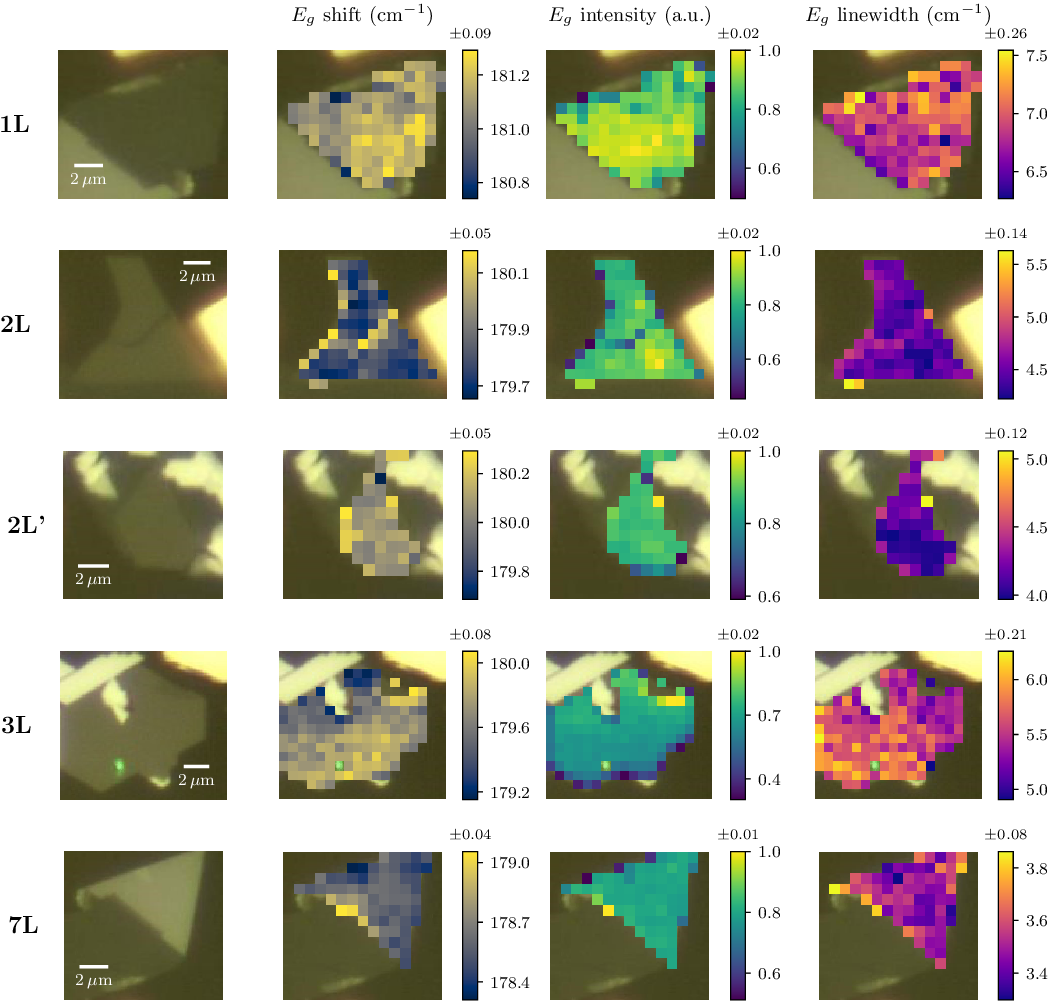}
\caption{Raman spectroscopy mappings of 1L, 2L, 2L', 3L and 7L flakes. The $E_g$ mode shift, intensity and linewidth are extracted using the 4-components Lorentzian fit detailed in the main text. These quantities exhibit significant variations on the edges of the flakes, while they remain relatively uniform across the surface (within the fit standard deviation error, represented on top of each scale bar). For these measurements, a Raman Invia spectrometer (Renishaw) is used, with a $\sim 50\,\rm{\mu W}$ $532\,\rm{nm}$ laser, $30\,\rm{s}$ integration time and a $1800\,\rm{gr/mm}$ grating.
  \label{fig:ramanMaps}}
\end{center}
\end{figure*}

\begin{figure*}[h!]
\begin{center}
\includegraphics[width=5in]{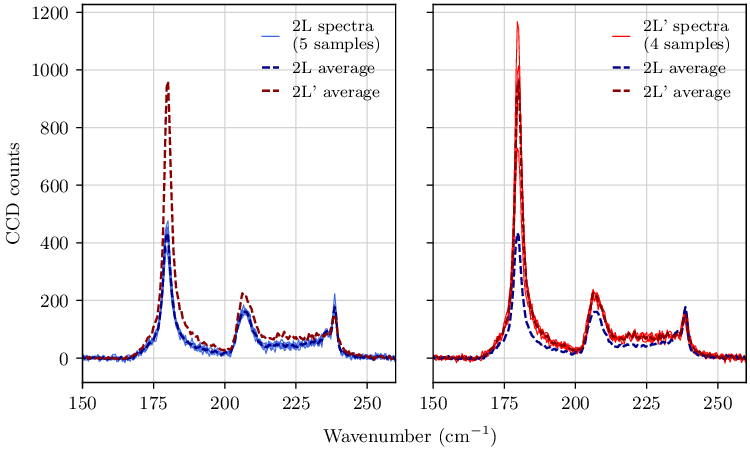}
\caption{Raman signatures of bilayer flakes, sorted as 2L (left) and 2L' (right). In dashed lines are represented the mean values of the 2L and 2L' spectra, highlighting that all bilayer samples can indeed be classified into these two categories.}
\label{fig:ramanBilayers}
\end{center}
\end{figure*}

\begin{figure*}[h!]
\begin{center}
\includegraphics[width=7in]{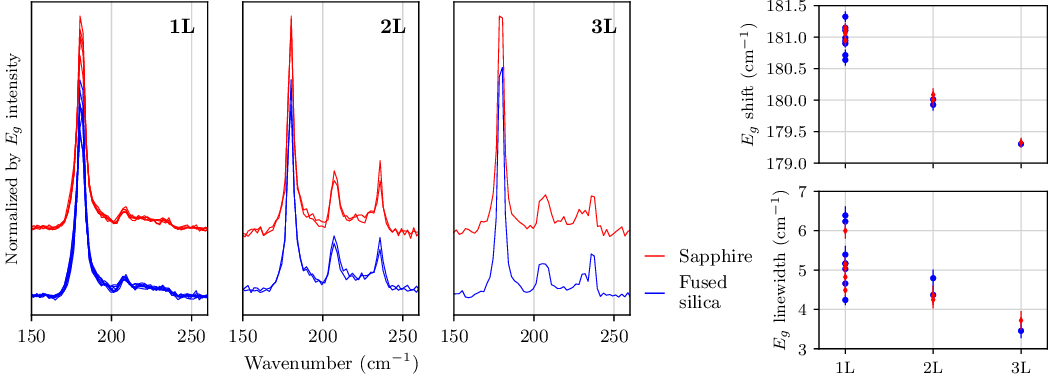}
\caption{Raman spectroscopy of 1L, 2L and 3L flakes on crystalline sapphire and on fused silica substrates. (left) Scattered Raman intensity, normalized by the $E_g$ mode integrated intensity, and the (right) $E_g$ mode shift and linewidth. A total of 8 flakes deposited on sapphire and 11 flakes deposited on fused silica have been inspected, respectively in red and blue. Raman signatures, $E_g$ shifts and linewidths appear to be very similar. For these measurements, a Raman Invia spectrometer (Renishaw) is used, with a $\sim 50\,\rm{\mu W}$ $532\,\rm{nm}$ laser, $150\,\rm{s}$ integration time and a $1800\,\rm{gr/mm}$ grating.}
\label{fig:ramanSapphire}
\end{center}
\end{figure*}

\begin{figure*}[h!]
\begin{center}
\includegraphics[width=4.5in]{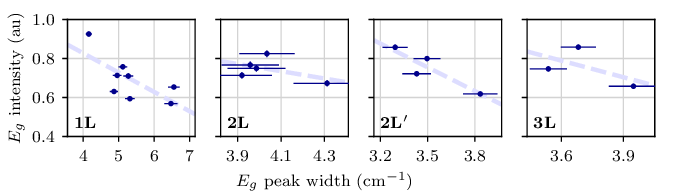}
\end{center}
\caption{$E_g$ peak intensity correlation with its linewidth, for $\rm{1L}$, $\rm{2L}$, $\rm{3L}$ -- and $\rm{2L'}$ bilayer flakes conjectured to originate from AB staking. The dashed lines are fitted, indicating the global trend. Each datapoint is originating from a different flake.
  \label{fig:EgWidthInt}}
\end{figure*}
